\documentclass[
    aps,
    prd,
    reprint,
    superscriptaddress,
    nolongbibliography,
    nobibnotes,
    nofootinbib,
]{revtex4-2}

\usepackage[dvipsnames, usenames]{xcolor}
\definecolor{linkcolor}{rgb}{0.1, 0.5, 0.7}

\usepackage[
colorlinks=true,
linkcolor=linkcolor,
citecolor=linkcolor,
filecolor=linkcolor,
urlcolor=linkcolor,
pdfusetitle,
]{hyperref}

\usepackage{amssymb}
\usepackage{amsmath}
\usepackage{physics}
\usepackage{orcidlink}
\usepackage{xspace}
\usepackage{multirow}

\newcommand{\comment}[1]{}

\newcommand{\gwTenElevenLR}{\ensuremath 4.1}
\newcommand{\gwElevenTenLR}{\ensuremath 0.8}

\newcommand{\CMCpctLowMassAstro}{\ensuremath 78}
\newcommand{\CMCpctLowMassDet}{\ensuremath 59}

\newcommand{\chieff}{\ensuremath{\chi_\mathrm{eff}}\xspace}
\newcommand{\chip}{\ensuremath{\chi_\mathrm{p}}\xspace}

\newcommand{\LRp}{LR\textsubscript{p}\xspace}
\newcommand{\LRj}{LR\textsubscript{joint}\xspace}

\newcommand{\ligo}{\affiliation{LIGO Laboratory, Massachusetts Institute of Technology, Cambridge, MA 02139, USA}}
\newcommand{\mki}{\affiliation{Kavli Institute for Astrophysics and Space Research, Massachusetts Institute of Technology, Cambridge, MA 02139, USA}}
\renewcommand{\mit}{\affiliation{Department of Physics, Massachusetts Institute of Technology, Cambridge, MA 02139, USA}}
\newcommand{\adler}{\affiliation{Adler Planetarium,1300 South DuSable Lake Shore Drive, Chicago, IL, 60605, USA}}
\newcommand{\ciera}{\affiliation{Center for Interdisciplinary Explorationand Research in Astrophysics (CIERA), Northwestern University, 1800 Sherman Avenue, Evanston, IL, 60201, USA}}

\begin{document}

\title{Surprisingly normal? GW241011 and GW241110 in the context of\\hierarchical merger classification}

\author{Cailin Plunkett\,\orcidlink{0000-0002-1144-6708}}\email{caplunk@mit.edu}
\ligo\mki\mit

\author{Michael Zevin\,\orcidlink{0000-0002-0147-0835}}\adler\ciera

\author{Salvatore Vitale\,\orcidlink{0000-0003-2700-0767}}
\ligo\mki\mit

\date{\today}

\begin{abstract}

Uncovering the origins of merging binary black holes is a key goal in gravitational-wave astrophysics, but identifying the formation channel of a particular binary is complicated by imprecise measurements and uncertain theoretical predictions. While classification often involves computing metrics from an event's inferred properties, measurement precision varies across parameter space. Using the perhaps-surprising low masses of the hierarchical merger candidates GW241011 and GW241110 as a backdrop, we consider the \textit{classifiable} distribution, which steps beyond detectability to assess what events may be correctly characterized as part of a population. Interpreting classification outcomes, and assessing the falsifiability of theoretical model predictions, are benefit by understanding the regions of parameter space that \textit{a priori} enable---and do not enable---confident measurement. For hierarchical mergers, we demonstrate that mass ratio is the largest intrinsic driver of identification, while orbital inclination and signal-to-noise ratio are the key extrinsic factors. Further, we show GW241011 and GW241110 achieve classification metrics consistent with similar simulated hierarchical mergers. Model selection and classification require the context of measurement capacity, especially as catalogs offer growing numbers of events with seemingly unusual properties.

\end{abstract}

\maketitle

\section{Introduction}

The gravitational-wave (GW) events GW241011 and GW241110 \citep{gw1011disc} reported by the LIGO--Virgo--KAGRA collaboration (LVK) \cite{LIGOScientific:2014pky, Capote:2024rmo, LIGO:2024kkz, VIRGO:2014yos, KAGRA:2020tym}
bear distinct spin magnitude and tilt signatures suggestive of hierarchical assembly. They possess remarkable spin properties opposite in nature, with GW241011 having the highest lower limit on component spin magnitude and GW241110 the lowest upper limit on spin-orbit misalignment of significant binary black hole (BBH) merger candidates reported by the LVK at the time of discovery \citep{gw1011disc, gwtc4pop, gwtc5pop}. Their mass asymmetry, high spin magnitudes, and spin-orbit misalignments are at odds with predictions from isolated binary evolution, which prefers small, aligned primary spins with mass ratios $q\gtrsim 0.8$~\cite{Kalogera:1999tq, Heger:2004qp, Belczynski:2016obo, Zaldarriaga:2017qkw, Qin:2018vaa, Gerosa:2018wbw, Fuller:2019sxi, Bavera:2020inc, Zevin:2022wrw}.\footnote{See Refs.~\cite{Popa:2025dpz, Gottlieb:2025ugy, Croon:2025gol, Bartos:2025pkv} for pathways to achieve high spins $\chi\gtrsim0.8$ from isolated evolution in the context of explaining GW231123.} Their properties are readily explainable, on the other hand, under the hierarchical hypothesis, in which the primary black hole was itself the product of a previous merger~\cite{Miller:2008yw, Rodriguez:2019huv, Doctor:2019ruh, Gerosa:2021mno, Mahapatra:2024qsy, Sedda:2023qlx}.

BBH merger remnants inherit the angular momenta of their progenitors, resulting in spin magnitudes peaked at $\chi\approx 0.7$ with small dispersion based on the progenitors' spins and mass ratio~\cite{Pretorius:2005gq, Buonanno:2007sv, Tichy:2008du, Rezzolla:2007rz, Fishbach:2017dwv}. Participation in subsequent mergers requires dense environments with high escape velocities to retain recoiling remnants~\cite{Fitchett:1983qzq, Favata:2004wz, Schnittman:2007sn, Gonzalez:2006md}. A \textit{hierarchical merger} between a first-generation (1G) and remnant second-generation (2G) black hole is distinguished by its primary spin of $\sim 0.7$, typical $\sim$1:2 mass ratio, and isotropic spin-tilt from dynamical assembly~\cite{OLeary:2005vqo}.

%%%%%%%%%%%%%%%%%%%%%%%%%%%%%%%%%%%%%%

The relatively low masses ($\sim15\text{--}20M_\odot$) of both GW241011 and GW241110 may come as a surprise given that hierarchical formation is frequently invoked as a mechanism to produce massive black holes. Hierarchical mergers are often cited as candidates to populate the upper mass gap: the purported dearth of black holes between $\sim50\text{--}130\ M_\odot$ due to pair-instability supernovae \cite{Barkat:1967zz, Heger:2002by, Belczynski:2016jno, Farmer:2019jed, Woosley:2021xba, Gerosa:2021mno}. Indeed, a hierarchical origin has been used to interpret high-mass, high-spin BBHs such as GW190521 and GW231123 \cite{LIGOScientific:2020iuh, Kimball:2020qyd,  Paiella:2025qld, Li:2025pyo, Li:2025fnf}. Lower-mass mergers like GW190412 have also received hierarchical interpretations, though typically with inconclusive evidence~\cite{Gerosa:2020bjb, Rodriguez:2020viw, Kimball:2020opk}. While the prevailing discussion centers on high-mass mergers, cluster simulations show hierarchical mergers may peak at primary masses $m_1\lesssim20M_\odot$, especially in higher-metallicity environments ($Z\gtrsim 0.1\ Z_\odot$)~\cite{Mapelli:2021gyv, Chatterjee:2016thb, Kremer:2019iul, Ye:2025ano, Agrawal:2026eqa, Vaccaro:2026nkp, OShaughnessy:2026dsr, OShaughnessy:2026xrd}.

%%%%%%%%%%%%%%%%%%%%%%%%%%%%%%%%%%%%%%

In this paper, we study the interpretation of 
GW241011 and GW241110 as hierarchical mergers, centering on the question: is it surprising that the most confidently hierarchical black-hole mergers are both low-mass? In the context of observations, the expected distribution is modified by the key word \textit{confidently}. Neither the astrophysical population nor the detectable subset is sufficient; we instead require the \textit{classifiable} population: those we can successfully identify as hierarchical. In particular, we ask whether high-mass mergers are systematically missed in hierarchical classification due to poorer spin measurements~\cite[e.g.,][]{Biscoveanu:2021nvg}.

No simple criterion enables perfect classification of hierarchical mergers. Theoretical predictions for different formation channels overlap and measurements are imprecise. Notably, Ref.~\cite{gw1011disc} did not compute quantitative metrics for GW241011 and GW241110, instead proposing hierarchical assembly as a more natural explanation than isolated evolution given the observed mass ratios and spins. Ref.~\cite{Payne:2024ywe} introduced a classification metric in effective precessing spin, \chip~\cite{Schmidt:2014iyl}, finding only a small fraction of 1G+2G mergers are classifiable due to measurement uncertainty. Ref.~\cite{Mould:2023ift} proposed using the posterior odds between two models, as Bayesian model selection naturally accounts for parameter estimation uncertainty, offering a consistent solution for comparing an event to different formation theories. However, which merger properties drive classification outcomes remains minimally explored for the hierarchical case.

Spin measurement precision varies across black-hole parameter space~\cite{Lang:2006bsg, Raymond:2009cv, Cho:2012ed, Baird:2012cu, Vitale:2014mka, Pankow:2016udj, Vitale:2016avz, Ng:2018neg, Chatziioannou:2018wqx, Shaik:2019dym, Fairhurst:2019srr, Pratten:2020igi, Green:2020ptm, Biscoveanu:2021nvg, Krishnendu:2021cyi, Biscoveanu:2026ikx}. For instance, effective inspiral spin \cite[\chieff, ][]{Racine:2008qv, Ajith:2009bn} measurements degrade with mass due to fewer inspiral cycles~\cite{Vitale:2016avz, Ng:2018neg} and component spins are better resolved at unequal mass ratios~\cite{Cho:2012ed, Baird:2012cu, Vitale:2014mka, Pratten:2020igi, Green:2020ptm}. It is critical to identify regions of parameter space where we fail to classify hierarchical mergers due to poor spin measurability as a major question for an astrophysical model is whether its predictions are \textit{falsifiable}. Even if a model produces detectable events, they may occupy a region that does not admit sufficiently precise measurements to distinguish formation channels. 

Using software injections and an astrophysically-informed simulated catalog, we investigate how hierarchical classification ability changes across parameter space. We find primary mass has minimal impact on classification; instead, mass ratio (important for precession modulation and breaking component label degeneracy) and redshift (which correlates with signal-to-noise ratio (SNR) and inclination angle) are the strongest predictors. We show GW241011 and GW241110 have metrics consistent with those expected from comparable simulated hierarchical mergers.

\section{Simulations of hierarchical mergers}\label{sec:pe-setup}

\subsection{Simulating mergers in a grid}
\label{ssec:gridpe-setup}

We systematically study how merger properties impact classification with simulated hierarchical signals in a grid across parameter space. For all ``grid" signals we set the primary spin magnitude $a_1 = 0.7$ and secondary $a_2 = 0$, as expected for 2G remnants produced by nonspinning 1G black holes. One at a time, we vary detector-frame primary mass $m_{1,\mathrm{det}}$, mass ratio $q$, and primary spin-tilt angle $\tau_1$ while holding other parameters constant. The simulated parameters are in Tab.~\ref{tab:gridPE}. We simulate mergers at two fixed values of network optimal SNR, 12 and 30, and adjust merger distance to maintain the chosen SNR.

\begin{table}[h]
\centering
\begin{tabular}{l|ll}
\hline \hline
Parameter & Values & \\
\hline
$m_{1,\mathrm{det}}\,[M_\odot]$ & 10, 20, 30, 50, 80, 120 & \\
$q$ & 0.25, (0.33), 0.5, 0.75, 1 & \\
$\tau_1\,[^\circ]$ & 0, 30, 50, 90, 120 & \\
$\theta_\mathrm{JN}\,[^\circ]$ & 30, (80) &\\
SNR & 12, 30 & \\
\hline \hline
\end{tabular}
\caption{Parameters and values of simulated hierarchical mergers used to investigate the dependence of classification on signal parameters. Values in parentheses are only covered for the SNR 12 injections. }
\label{tab:gridPE}
\end{table}

The angle between the total angular momentum and line of sight, $\theta_\mathrm{JN}$ is known to impact spin measurements \cite{Vitale:2014mka}. Systems viewed more edge-on ($\theta_\mathrm{JN}\sim\pi/2$) have typically better-measured spins because the precession of the spin vectors around the total angular momentum induces more variation in the waveform along the line of sight. However, such systems are harder to detect, with lower SNR compared to face-on. At SNR 12, we inject at $\theta_\mathrm{JN}=30^\circ$ and $80^\circ$: both near the detectability peak and a configuration more favorable for measuring spin. As we expect most high-SNR systems to be nearly face-on, we only inject at $\theta_\mathrm{JN}=30^\circ$ for SNR 30. At SNR 12, we consider one additional value of mass ratio, $q = 0.33$.

We select the sky location to be near-optimal for LIGO Hanford and LIGO Livingston, choosing directly overhead the midpoint of the great circle that connects the two observatories. We have verified the sky location does not change our conclusions with a smaller injection set at varied positions (see also Ref.~\cite{Vitale:2014mka}). We inject the signals into a three-detector LIGO--Virgo network in zero noise at the design sensitivity for the fourth observing run~\cite{LIGO-T2000012} (App.~\ref{app:detectability}). Further details of our parameter estimation (PE) configuration are given in App.~\ref{app:PE}.

\subsection{Simulating realistic stellar cluster mergers}\label{ssec:cmc-mergers}

To study a realistic population of hierarchical mergers, we employ the \textsc{Cluster Monte Carlo} \citep[CMC, ][]{Kremer:2019iul, Rodriguez:2021qhl} catalog, which consists of 148 $N$-body simulations of stellar clusters. These simulations cover a range of initial conditions, with 144 of the models tiling a hypercube in the number of stars ($N = \{2,4,8,16\}\times10^5$, corresponding to cluster masses $M = \{1.2,2.4,4.8,9.6\}\times10^5 M_\odot$), virial radius ($r_v = \{0.5,1,2,4\}$ pc), galactocentric distance ($r_g = \{2,8,20\}$ kpc), and metallicity ($Z = \{0.01, 0.1, 1\} \ Z_\odot$). Four additional simulations at $N = 3.2\times10^6, \ r_g = 20$ kpc, $r_v = \{1,2\}$ pc, and $Z = \{0.01,1\} \ Z_\odot$ round out the catalog. Each simulation tracks stellar evolution and interaction over the cluster's lifespan, from which the number and properties of all BBH mergers are collated. As the initial cluster conditions affect BBH properties, the astrophysical population distribution of dynamical BBHs thus depends on the formation history and properties of clusters. We couple these simulations to chosen weightings to create an astrophysical distribution of cluster mergers.

\subsubsection{Astrophysical weighting}\label{sssec:weighting}

To draw astrophysically-informed samples from the CMC catalog, we follow a similar procedure to Refs.~\citep{Fishbach:2023xws, Ye:2024ypm, Ye:2025ano}. We draw formation redshifts from the comoving cluster formation rate density $z_\mathrm{form}\sim \mathcal R_\mathrm{GC}(z)$. The cluster formation rate is chosen to follow the star formation rate, described by a Madau-Dickinson model~\cite{Madau:2016jbv}:
\begin{align}
    \mathcal R_\mathrm{GC} \propto \frac{(1+z)^\lambda}{1 + \left(\frac{1+z}{1+z_p}\right)^{\lambda+\kappa}},
\end{align} with $\lambda = 2.6, \ \kappa = 3.6,$ and $z_p = 2.2$. While cluster formation likely peaked before star formation~\cite{ElBadry2019}, we choose this formation rate to match Refs.~\cite{Ye:2024ypm, Ye:2025ano}.

Each simulation is weighted based on assumed distributions of cluster parameters. The weight is the probability of forming a cluster with parameters $(M, r_v, Z)$ at $z_\mathrm{form}$. As we have a discrete set of simulations at equal log-spaced intervals in parameter space, we integrate the weights over each bin such that the probability of cluster formation is $\Lambda \,p(\Lambda \mid  z_\mathrm{form})$. All binaries within each cluster inherit their cluster's weight if they merge before redshift $z=0$; otherwise, they are given zero weight. Each binary's time of merger is $t_{\mathrm{merge},i} = t(z_\mathrm{form}) - t_{\mathrm{delay},i}$, where $t(z_\mathrm{form})$ is the lookback time at $z_\mathrm{form}$ and $t_{\mathrm{delay},i}$ is the time delay between cluster formation and BBH merger. The weight for each binary is then 
\begin{align}\label{eq:weights}
    w_i &=  p(M_i) M_i \ p(r_{v,i}) r_{v,i} \ p(Z_i\mid z_\mathrm{form}) Z_i \ \Theta(t_{\mathrm{merge},i}),
\end{align} where $\Theta$ is a Heaviside function. As in Refs.~\cite{Fishbach:2023xws, Ye:2024ypm, Ye:2025ano}, we do not weight by galactocentric distance.

We sample one binary according to these weights for each drawn formation redshift. This produces an astrophysical distribution of cluster mergers in the comoving source frame. To create a catalog in the detector frame, we account for the geometric and time-delay factors from converting to a detector-frame rate from a comoving source-frame rate density. We resample the drawn binaries according to the volume at $z_{\mathrm{merge}, i} = z(t_{\mathrm{merge},i})$ and conversion to detector-frame time using
\begin{align}\label{eq:weights2}
    w_{2,i} &=  \frac{\dd V_c}{\dd z_{\mathrm{merge},i}} \ \frac{1}{1+z_{\mathrm{merge},i}}.
\end{align}

For the spectra of cluster masses, radii, and metallicities in Eq.~\ref{eq:weights}, we adopt the same distributions as Ref.~\cite{Ye:2024ypm}. Cluster masses are described by a Schechter function~\cite{Schechter:1976iz},
\begin{align}
    p(M)\propto \left(\frac{M}{M_S}\right)^\beta\exp\left(-\frac{M}{M_S}\right),
\end{align}
with $\beta = -2$ and $M_S = 10^{6.26}\ M_\odot$ \cite{PortegiesZwart:2010cly, Antonini:2020xnd}. We assume cluster radii follow a Gaussian
\begin{align}
    p(r_v)\propto\exp\left(-\frac{(r_v-\mu_r)^2}{2\sigma_r^2}\right),
\end{align} with $\mu_r = 2$ pc and $\sigma_r = 2$ pc \cite{Fishbach:2023xws}
Finally, metallicity is assumed to be distributed lognormally with a redshift-dependent mean,
\begin{align}
    p(Z\mid z)\propto \frac1Z\exp\left(-\frac{(\log_{10}Z-\mu_Z(z))^2}{2\sigma_Z^2}\right),
\end{align} where $\sigma_Z=0.5$ and $\mu_Z$ is given by \cite{Madau:2016jbv}
\begin{align}\label{eq:mu_z}
    \log_{10}\langle Z/Z_\odot\rangle(z) = 0.153 - 0.074z^{1.34}.
\end{align}

We draw a bootstrapped sample of $n = 500,000$ binaries from the CMC catalog, of which roughly $\sim10\%$ are 1G+2G. We assess detectability using representative detector sensitivities~\cite{LIGOScientific:2025hdt, Virgo:2022ysc} then perform PE on the 755 detected 1G+2G binaries using the methods detailed in Apps.~\ref{app:detectability} and~\ref{app:PE}.\footnote{While Ref.~\cite{Payne:2024ywe} also simulated hierarchical mergers drawn from the CMC catalog, we opt to run our own PE using updated detection criteria, detector sensitivities, and astrophysical weights.}

\section{Classification metrics} \label{sec:metrics}

Classifying hierarchical mergers is a non-trivial task because their distinguishing features---high primary spin, spin-orbit misalignment, and unequal masses---can be difficult to infer from GW signals. Here we use the likelihood ratio (LR) statistic introduced in Ref.~\cite{Payne:2024ywe}. They observe first-generation and hierarchical mergers separate cleanly in precessing spin \chip, with most first-generation (hierarchical) mergers possessing $\chip<0.2$ ($\chip\geq0.2)$. Their LR quantifies the support above versus below a threshold in a spin parameter $\chi$ as
\begin{align}\label{eq:LR}
    \mathrm{LR}_{\chi\leq\chi_\mathrm{thres}}^{\chi>\chi_\mathrm{thres}} = \frac{\int_{\chi_\mathrm{thres}}^{\chi_\mathrm{U}}\mathcal L(d\mid\chi)\,\mathcal U(\chi_\mathrm{thres},\chi_\mathrm{U})\dd\chi}{\int_{\chi_\mathrm{L}}^{\chi_\mathrm{thres}}\mathcal L(d\mid\chi)\,\mathcal U(\chi_\mathrm{L},\chi_\mathrm{thres})\dd\chi},
\end{align} where $\mathcal L(d\mid\chi)$ is the marginal likelihood for data $d$ given a spin parameter $\chi$ and $\mathcal U$ denotes a uniform prior. Their simulated mergers inform both the spin parameter threshold $\chi_\mathrm{thres}=0.2$ in \chip and the LR classification threshold. They identify a weak threshold of $\ln\mathrm{LR}\geq3$ (successfully classifying 8\% of hierarchical mergers) and a strong threshold of 8 (classifying 2\% of mergers). We use $\ln\mathrm{LR}\geq3$ for classification.

While \chip can be a powerful differentiator of hierarchical and first-generation mergers, it also compresses the full seven-dimensional spin and mass ratio space to a single parameter. In practice, combining information across multiple dimensions is key to both successful identification and avoidance of false positives. We introduce a likelihood ratio in a multidimensional parameter space $\boldsymbol\theta$ as an alternate classification metric:
\begin{align}\label{eq:LR-joint}
    \mathrm{LR}_{\boldsymbol\theta\notin\mathcal{V}}^{\boldsymbol\theta\in\mathcal{V}} = \frac{\int_{\mathcal V} \mathcal L (d\mid\boldsymbol\theta)\,\mathcal U(\mathcal V)\,\dd\boldsymbol\theta}{\int_{\mathcal{V}^c}
      \mathcal{L}(d\mid\boldsymbol\theta)\,\mathcal U(\mathcal V^c)\,\dd\boldsymbol\theta},
\end{align}
where the volume $\mathcal V$ in the multidimensional domain is expected to enclose most hierarchical mergers and few first-generation ones, and $c$ denotes its complement. We call this metric the ``joint LR."

We focus on the space of $\{\chieff, \chip\}$, which may offer discrimination between formation channels with better precision than component spin parameters~\cite{Baibhav:2021qzw, Antonini:2024het, Plunkett:2026pxt}. To mimic the simplicity of the threshold in Ref.~\cite{Payne:2024ywe}, we define $\mathcal V^c$ as the box enclosing $\chieff\in(-0.25, 0.25)$ and $\chip\in(0,0.2)$. While the physical region of the joint \chieff--\chip space is $q$-dependent, we approximate the total prior volume by one half of the unit circle. We continue to use a classification threshold of $\ln\mathrm{LR}\geq3$.

We validate the joint metric using the PE from Ref.~\cite{Payne:2024ywe} available at Ref.~\cite{PayneZenodo}. Of their 1000 1G+1G simulated signals at O3 sensitivity, none have a joint $\ln\mathrm{LR}>3$. Roughly $\sim16\%$ of their simulated hierarchical mergers are classified via the joint metric, double those classified by \chip alone. Under the small natal spin hypothesis, the joint metric is both robust to false positives and achieves better classification rates than \chip alone.

For notational clarity, we call the \chip-only metric \LRp and the metric in the joint \chieff--\chip space \LRj. We report results on real events, before analyzing the grid PE in Sec.~\ref{sssec:grid-pe-results} and the CMC PE in Sec.~\ref{sssec:cmc-merger-results}. In App.~\ref{app:metrics}, we additionally show a component likelihood ratio in the joint space of $\{q,\,a_1,\,\cos\tau_1\}$.

\subsection{Results on real events}

We compute the likelihood ratio classification metrics on the 259 BBH candidates with false alarm rate $<1\,\mathrm{yr}^{-1}$ in the cumulative Fifth Gravitational-Wave Transient Catalog~\cite{gwtc5cat, gwtc5_pe_release_part1}. We use PE from the  \textsc{NRSur7dq4}~\cite{Varma:2019csw} waveform when available, followed by \textsc{IMRPhenomXPHM-SpinTaylor}~\cite{Colleoni:2024knd}, then \textsc{IMRPhenomXPHM}~\cite{Pratten:2020ceb}. Four events are classified as hierarchical under the weak threshold of ln \LRp$>3$: GW231123 ( ln \LRp$=9.7$), GW241011 (\gwTenElevenLR), GW241127\_061008 (4.0), and GW241225\_082815 (4.4). Ten events achieve ln \LRj$>3$, the highest being GW241011 (infinite, as its \chieff is well-constrained to $\sim0.5$), GW241113\_163507 (9.9), GW231123 (9.3), and GW190517 (6.6). Many of these events have been previously identified as hierarchical candidates~\cite{Kimball:2020qyd, Paiella:2025qld, Li:2025pyo, Li:2025fnf, Tong:2025xir, Plunkett:2026pxt, Alvarez-Lopez:2026ymo}.

In the remainder of this work, we specify to GW241011 and GW241110 as our real-event candidates because they were specifically highlighted by the LVK in Ref.~\cite{gw1011disc} as hierarchical candidates and they lie in the potentially unexpected low-mass range, though we find other events achieve similar and often higher classification metrics. GW241110 is not classified as hierarchical under either metric: its ln \LRp$=\gwElevenTenLR$, and  ln \LRj$=1.2$. However, this does not exclude the hierarchical hypothesis: only $8\%$ of the simulated hierarchical mergers in Ref.~\cite{Payne:2024ywe} surpass the weak threshold. We note Ref.~\cite{gw1011disc} presents PE results mixed across three waveforms. Using mixed samples, GW241011's ln \LRp drops to 2.8, driven by lower inferred precession by \textsc{IMRPhenomXO4a}~\cite{Thompson:2023ase}; its \LRj remains infinite. GW241110's metrics are mildly affected by mixing waveform samples, with ln \LRp$=0.9$ and ln \LRj$=1.0$.

\section{Are GW241011 and GW241110 surprising?}\label{sec:results}

\subsection{Detectable hierarchical mergers}

\begin{figure}[!ht]
\centering
\includegraphics[width=\columnwidth]{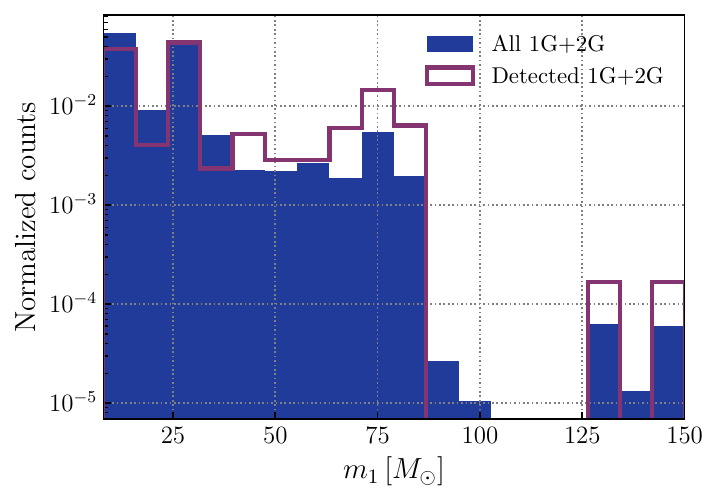}
\caption{The primary mass distribution for an astrophysical population of hierarchical mergers (blue, filled) and the detectable distribution (purple, step).}
\label{fig:cmc-all-vs-det}
\end{figure}

Before accounting for spin measurability, we first explore whether the detectable distribution of hierarchical mergers aligns with the properties of GW241011 and GW241110. Fig.~\ref{fig:cmc-all-vs-det} shows the primary mass distribution of the astrophysical hierarchical merger population (Sec.~\ref{ssec:cmc-mergers}) in blue and those that are detectable in purple. The sharp dropoff at $m_1\approx80M_\odot$ corresponds to roughly twice the assumed pair-instability limit for 1G black holes. The handful of 2G black holes above $80 M_\odot$ come from 1G black holes with masses in the pair-instability gap that formed from stellar collisions.

The selection function favors higher-mass events. While \CMCpctLowMassAstro\% of the astrophysical 1G+2G binaries have primary masses below $25\,M_\odot$, \CMCpctLowMassDet\% of the detectable mergers do. With a simple frequentist approach to the prevalence of lower-mass hierarchical mergers, we expect two consecutive low-mass mergers $\sim 35\%$ of the time. This basic figure implies it is reasonably likely for the first two detected hierarchical mergers to be below $25\,M_\odot$.

\begin{figure*}[!ht]
\centering
\includegraphics[width=0.9\textwidth]{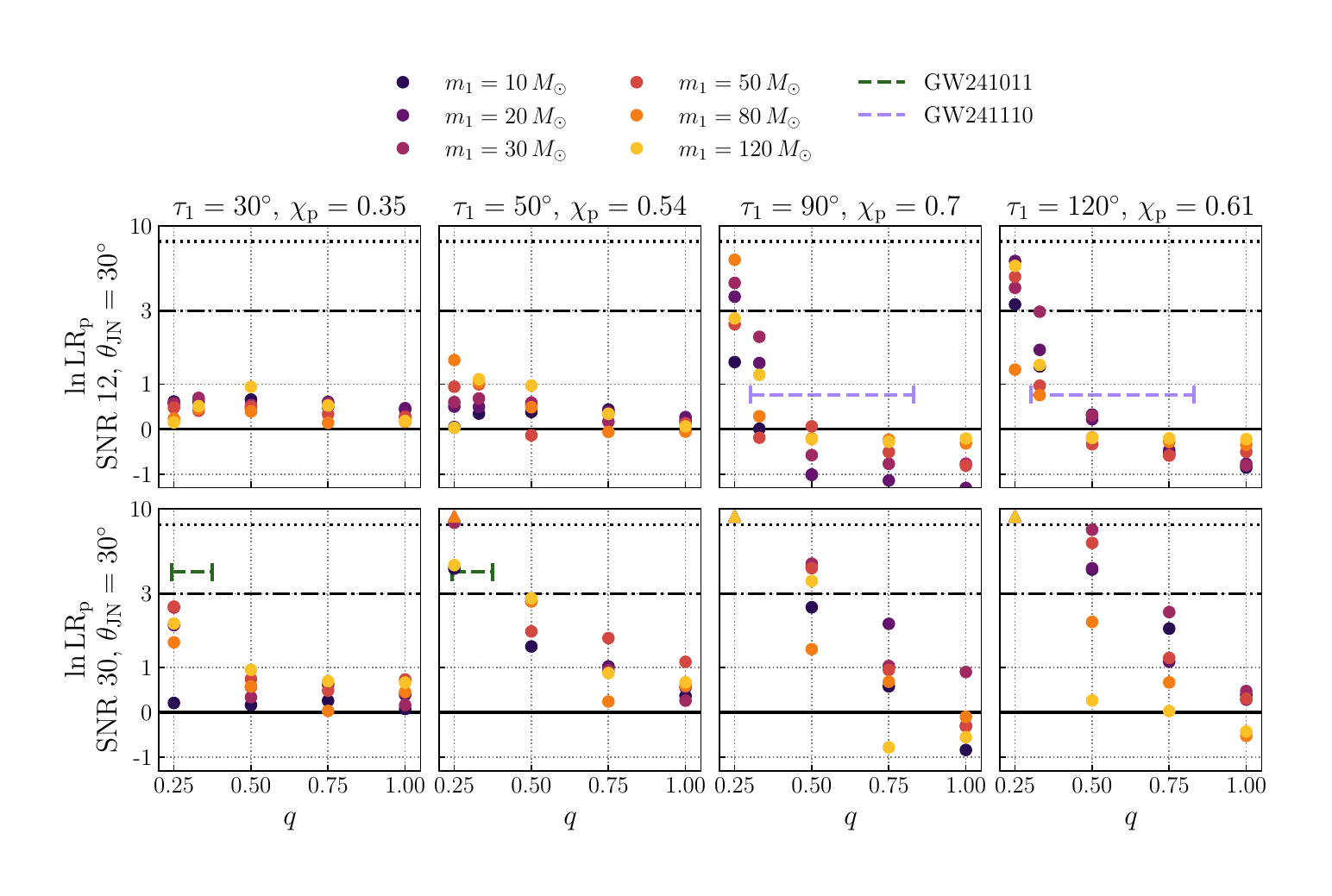}
\caption{The likelihood ratio statistic (Eq.~\ref{eq:LR}) for precessing spin \chip as a function of mass ratio for simulated 1G+2G systems with inclination angle $\theta_\mathrm{JN} = 30^\circ$. The top (bottom) row is at SNR 12 (30). Each column shows a different tilt angle and the colors represent different primary masses. 
The solid horizontal line at 0 denotes equal likelihood above and below the threshold of $\chip=0.2$, while the dash-dotted and dotted lines at 3 and 8 are the weak and strong thresholds of Ref.~\cite{Payne:2024ywe}. Triangles mark simulated events whose $\ln\mathrm{LR}$s are above the plot limit of 10.
The LRs for GW241011 and GW241110 are plotted as dashed dark green and light purple lines, respectively, that span the 90\% credible regions of the events' mass ratios. For a more realistic comparison, we only show the events' metrics on panels with tilts and SNRs similar to their inferred properties.}
\label{fig:PE_grid_LR}
\end{figure*}

\subsection{Classifiable hierarchical mergers}

We now turn to which hierarchical systems we can identify based on their PE posteriors. We apply the likelihood ratio criteria introduced in Sec.~\ref{sec:metrics} to the grid mergers and realistic CMC population.

\subsubsection{Grid mergers}
\label{sssec:grid-pe-results}

Our systematic exploration of hierarchical mergers show that at a given SNR and $\theta_\mathrm{JN}$, mass ratio is the largest predictor of hierarchical classification, which is consistent with the expectation that precession measurement improves with mass asymmetry~\cite{Green:2020ptm, Cho:2012ed, Pratten:2020igi}. Fig.~\ref{fig:PE_grid_LR} shows LR versus mass ratio for $\theta_\mathrm{JN} = 30^\circ$ systems at SNR 12 (top) and 30 (bottom), with columns showing increasing tilt and colors primary mass. LR strongly decreases toward more equal masses. At $\tau_1=90^\circ$, there is a slight correlation between LR and primary mass, but the effect is neither monotonic nor general. The nearly-aligned systems with $\tau_1=30^\circ$ have a relatively low $\chip=0.35$, such that even the best-measured injections have significant support below $\chi_\mathrm{thres}=0.2$ and thus low LRs. We obtain comparable precision at SNR 30 for the other tilt angles, with increasing variability at larger tilts.

At SNR 12 and  $\theta_\mathrm{JN}=30^\circ$, the only events classified have $q=0.25$ and $\tau_1\geq90^\circ$. The LRs at $q\geq0.5$ systematically decrease as tilt increases, which may arise from the shorter inspiral leading to more poorly measured spins. At the more edge-on inclination, $\theta_\mathrm{JN}=80^\circ$, LRs are typically higher though have larger variability.

\begin{figure*}[!ht]
\centering
\includegraphics[width=0.9\textwidth]{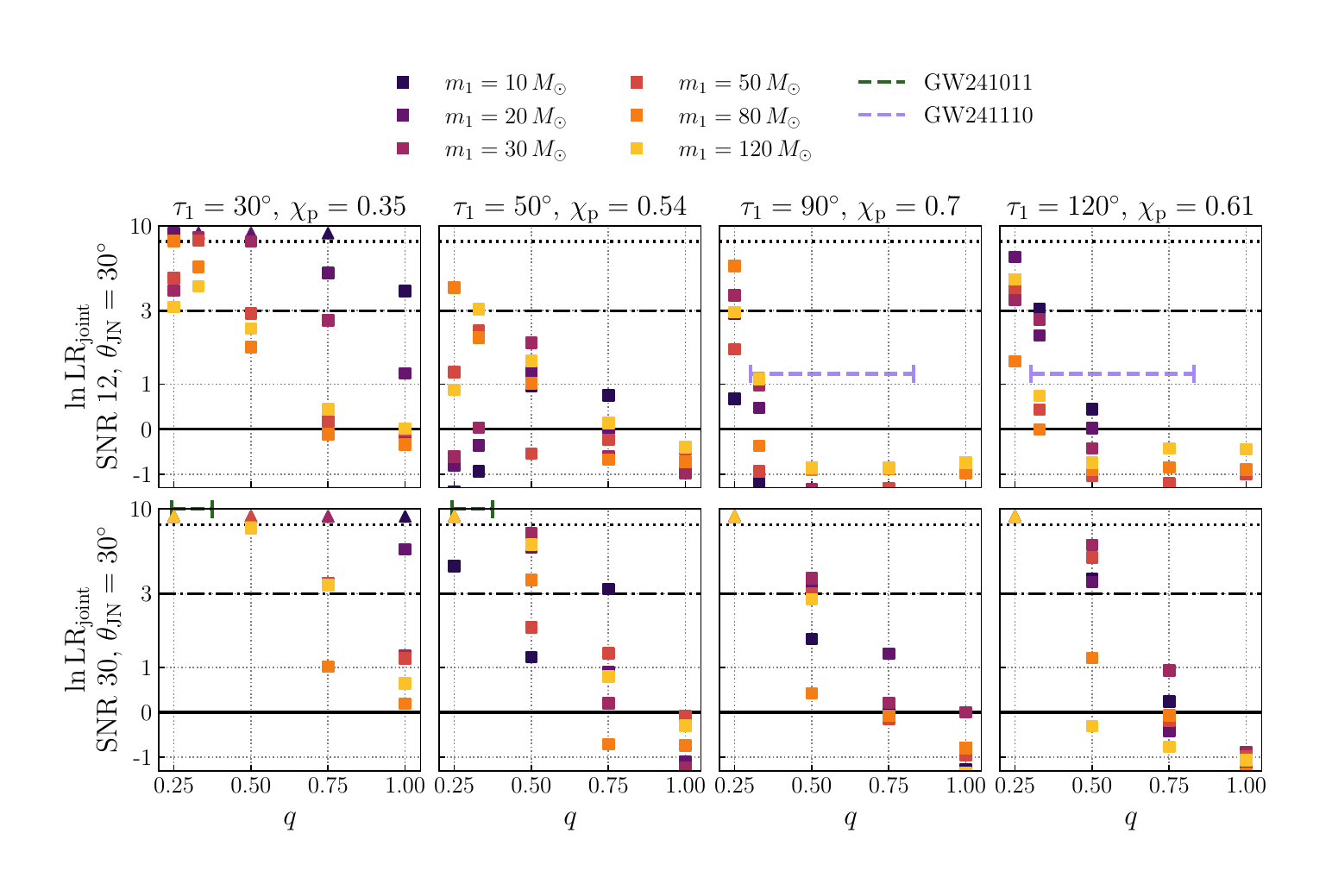}
\caption{As in Fig.~\ref{fig:PE_grid_LR} but for the likelihood ratio metric in the joint space of \chieff and \chip (Eq.~\ref{eq:LR-joint}).}
\label{fig:PE_grid_LR_joint}
\end{figure*}

The trends with the \chieff--\chip joint metric are similar, except aligned mergers (with high true \chieff) are better captured, shown in Fig.~\ref{fig:PE_grid_LR_joint}. Mass ratio continues to be a strong driver of classification. Since high, positive \chieff can be measured more precisely than negative \chieff~\cite{Ng:2018neg}, the $\tau_1 = 30^\circ$ mergers are well-captured by the joint LR and tilt angle becomes less critical for classification.

Waveform modulation due to precession increases with larger opening angle between the total and orbital angular momenta, which in turn scales with $q$~\cite{Green:2020ptm, Pratten:2020igi}. The precision of \chip measurement, and therefore classification probability, thus worsens toward equal masses. Label confusion also contributes to imprecise measurements of component spin parameters toward equal mass~\cite{Lang:2006bsg, Vitale:2014mka, Pankow:2016udj, Biscoveanu:2020are}. When the components are not distinguishable in mass, the nonspinning secondary can contaminate measurements of the primary's rapid spin. Even at $q=0.5$, it is rare to confidently constrain $q$ away from unity at typical SNRs ($\lesssim 15$), so label confusion thus contributes to weaker constraints on primary spin and tilt angle.

Measurements of inspiral parameters, such as chirp mass and \chieff, worsen with mass due to the smaller number of inspiral cycles in-band before merger~\cite{Vitale:2014mka, Shaik:2019dym}. However, we find those of the constituents of \chieff ($q$, $a_1$, and $\tau_1$) do not monotonically broaden with mass; only in the full induced distribution of \chieff is broadening significant. Important for classification, the posteriors on \chip most clearly degrade with $q$, not mass, consistent with Refs.~\cite{Green:2020ptm, Biscoveanu:2021nvg}. We further study component parameter measurement behavior over mass in App.~\ref{app:mass}.

The $\ln\mathrm{LR}$s for GW241011 (dark green) and GW241110 (light purple) are depicted in Fig.~\ref{fig:PE_grid_LR} as dashed lines that span the 90\% credible regions of their mass ratio posteriors. We only show GW241011 (GW241110) on the panels with SNR 30, $\tau_1\in\{30^\circ, 50^\circ\}$ (SNR 12, $\tau_1\in\{90^\circ, 120^\circ\}$), as these are the injected properties most consistent with the event's inferred values. While neither event has $\ln\mathrm{LR}>3$, their $\ln\mathrm{LR}$s are within the range expected for simulated hierarchical mergers with similar properties. This said, the injections are performed under idealized conditions and we do not compare to an injected population of first-generation mergers. Relaxing the idealizations (e.g. zero noise, detector design sensitivity) would likely reduce the mean LR and add variation. While our results do not necessarily increase the credence for hierarchical origins of GW241011 and GW241110, they demonstrate qualitative consistency between these events and comparable hierarchical mergers.

\subsubsection{CMC mergers}
\label{sssec:cmc-merger-results}

\begin{figure*}[!ht]
\centering
\includegraphics[width=0.9\textwidth]{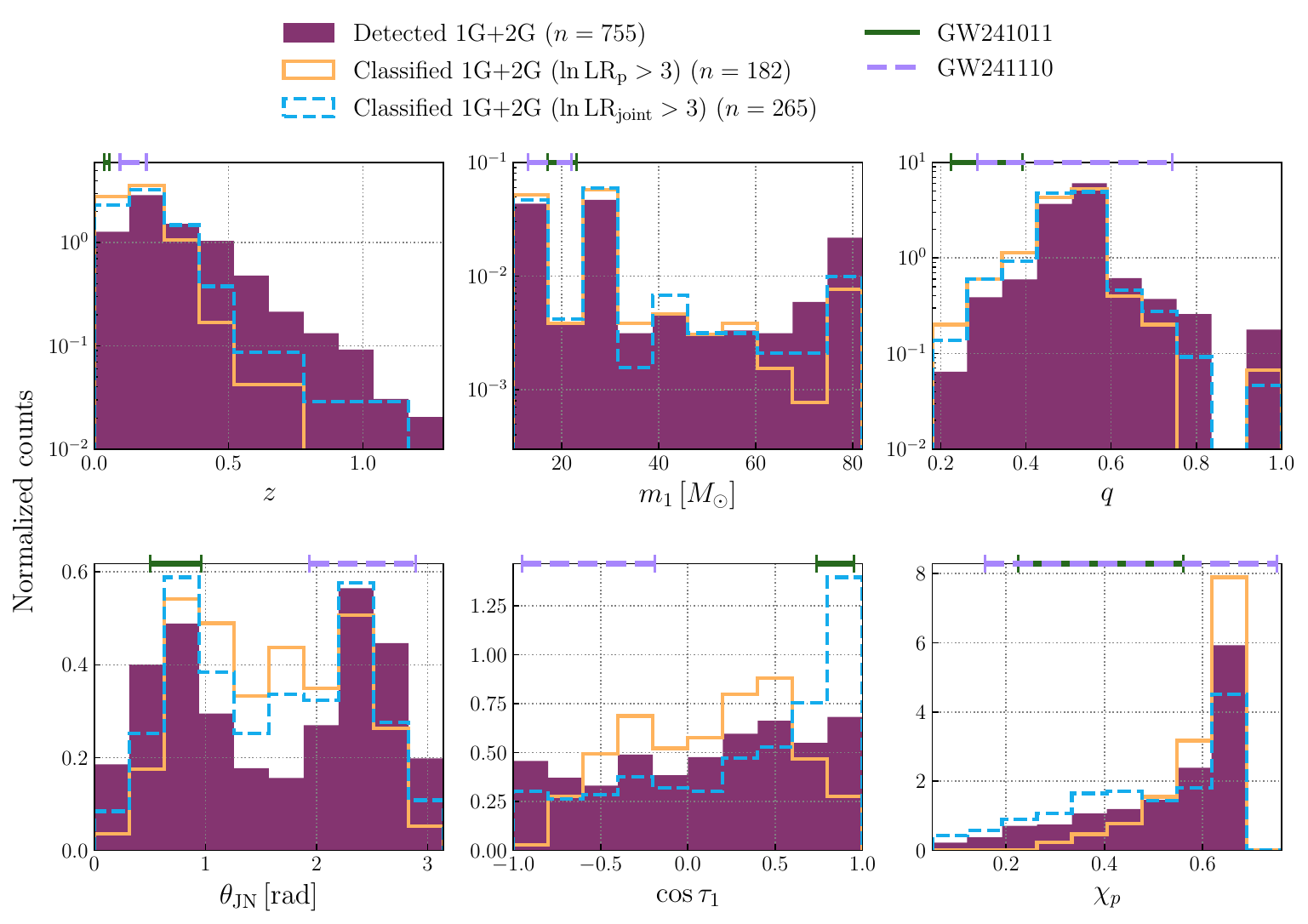}
\caption{The detected (filled magenta) and classified (step orange, dashed step blue) distributions of hierarchical mergers, based on astrophysical simulations, for various binary parameters. Systems are considered classified if their $\ln\mathrm{LR}$ statistic is $>3$; the orange curve classifies with \chip alone (Eq.~\ref{eq:LR}), while the dashed blue curve uses a joint metric in \chieff and \chip (Eq.~\ref{eq:LR-joint}). The upper axis limits are overlaid with the 90\% credible intervals for GW241011 (solid dark green) and GW241110 (dashed light purple).}
\label{fig:cmc-pe-det-vs-class}
\end{figure*}

Having systematically explored how binary parameters impact classification, we now consider the astrophysically-realistic population using the CMC catalog. Fig.~\ref{fig:cmc-pe-det-vs-class} shows the marginal parameter distributions of detected 1G+2G mergers as filled magenta histograms. Overlaid in orange (dashed blue) are the events classified according to $\ln\mathrm{LR_p}>3$ ($\ln\mathrm{LR_{joint}}>3$). On the top of each axis, we provide the 90\% credible regions for GW241011 and GW241110.

\begin{figure*}[!ht]
\centering
\includegraphics[width=0.9\textwidth]{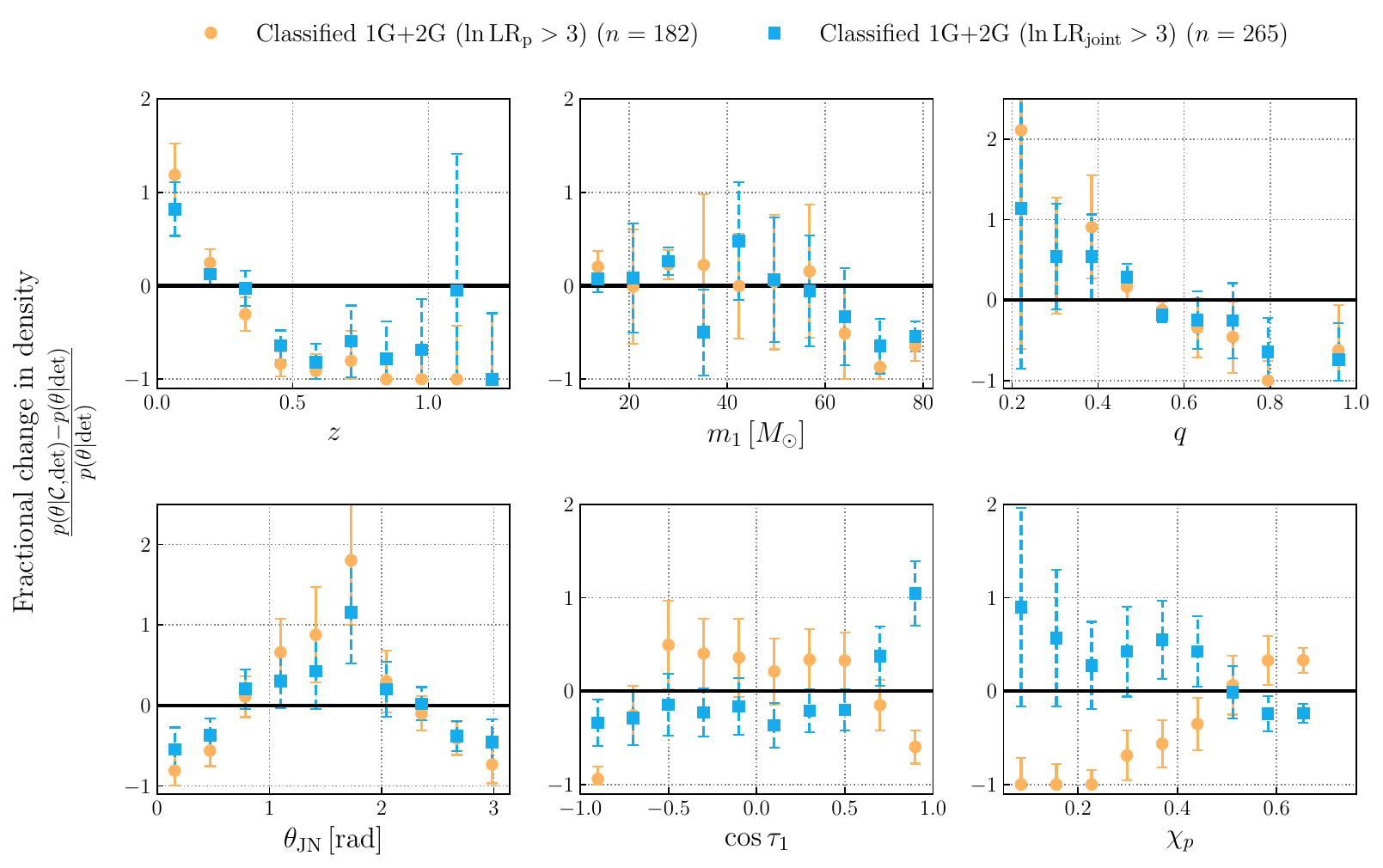}
\caption{The fractional change in density between the detected and classified populations of simulated hierarchical mergers at binned values in parameter space. Solid black lines at zero indicate no difference in density. Orange squares use the \chip metric; blue circles classify using the joint \chieff--\chip. Errorbars indicate 1-$\sigma$ binomial uncertainty, with lower bounds clipped to the minimum physical value of $-1$.}
\label{fig:cmc-pe-frac-change}
\end{figure*}

Where the orange and blue histograms are higher (lower) than the magenta indicates a region of parameter space where it is easier (harder) to classify a merger as hierarchical. The observed trends are similar to those identified with the grid PE, yet display additional features due to correlations between parameters in the astrophysical and detectable distributions. We also visualize these trends by computing the fractional change in density between the detected and classified population for a parameter $\theta$,
\begin{align}
    \Delta p(\theta) = \frac{p(\theta\mid\mathcal C,\mathrm{det})-p(\theta\mid\mathrm{det})}{p(\theta\mid\mathrm{det})},
\end{align} where $\mathcal C$ indicates an event was classified. The fractional density differences in bins across parameter space are shown in Fig.~\ref{fig:cmc-pe-frac-change} with 1-$\sigma$ binomial error bars. Values above zero indicate regions where mergers are preferentially classified. We first discuss the total classification rate before expounding on the six shown parameters.

\textbf{\textit{Classification rate}}: Overall, 182 (24\%) mergers are classified by \chip, while 265 mergers (35\%) are classified by the joint \chieff--\chip metric. The additional mergers found by the joint metric tend to be nearly-aligned mergers with low \chip but high \chieff. We further explore the difference in efficacy between \LRp and \LRj in App.~\ref{app:metrics}. 

Our 24\% classification rate is a factor of three higher than found in Ref.~\cite{Payne:2024ywe}. The discrepancy likely arises from differences in the adopted detector sensitivity and astrophysical weighting. Improvements in detector sensitivity from the third (O3) to first part of the fourth observing run enhance the SNRs of our simulated events, tightening parameter measurements. Second, Ref.~\cite{Payne:2024ywe} used a cluster formation rate~\cite{ElBadry2019} that peaks at a higher redshift ($z\sim 4$) than our chosen formation rate. This drives our higher detection rate and SNRs, contributing to better classification. The peak of formation for clusters that produce hierarchical mergers may lie between these two values, as many clusters are disrupted early in their lifetimes~\cite{Joschko2024}. The cluster formation history primarily affects the overall detection rate. While redshift is correlated with cluster metallicity and thus BBH mass in our weighting scheme, we find these two functional forms produce similar detectable distributions, so we thus expect similar conclusions on classification trends.

\textbf{\textit{Redshift and inclination}}: With both metrics, classifiability declines more steeply with redshift than detectability, as seen in the upper left panels of Figs.~\ref{fig:cmc-pe-det-vs-class} and~\ref{fig:cmc-pe-frac-change}. The fraction of classified binaries with $z\lesssim 0.1$ is twice that of detected binaries with $z\lesssim 0.1$. With \chip only, no merger is classified above $z\approx0.6$. While this is partially an SNR effect---closer systems typically have higher SNRs and thus better-measured parameters---SNR alone does not explain the drop in classification rate (see App.~\ref{app:metrics}, Fig.~\ref{fig:cmc-pe-snr}). The inclination selection bias drives much of the additional decline: farther systems are preferentially face-on because it boosts SNR, but face-on inclination worsens precession measurement.

The joint metric is less sensitive to redshift than the \chip-only metric because \chieff measurement does not require the orbital plane modulation that makes \chip sensitive to $\theta_\mathrm{JN}$. Adding \chieff as a classification dimension therefore reduces sensitivity to inclination. As the lower left panel of Fig.~\ref{fig:cmc-pe-frac-change} indicates, \LRj misses fewer face-on events and is less dominated by edge-on signals. By correlation, \LRj is thus less dependent on redshift: \chieff can be well-measured even at high redshifts where most events are face-on. Indeed, for nearly edge-on events, the two metrics show indistinguishable redshift dependence. For face-on events, in contrast, \chip-only classification drops more steeply (see App.~\ref{app:metrics}, Fig.~\ref{fig:compare-redshift-dependence}). This confirms that the difference in redshift dependence arises because redshift and inclination are correlated in detected events, with the joint metric succeeding even where inclination suppresses precession modulation because inspirals can still encode \chieff.

\textbf{\textit{Primary mass}}: There is a slight drop in classifiability at the highest source-frame primary masses, which is intriguing as a contributing factor for the low masses of the observed events. This is a consequence of the catalog's positive correlation between mass and redshift, which is induced by selection effects and the adopted evolution of cluster metallicity across cosmic time. No trend with mass (detector- or source-frame) was observed in the grid injections, which have uncorrelated parameters by construction. Indeed, regression analyses (Sec.~\ref{sssec:logistic-regression}) indicate that mass's apparent effect lies in its correlation with redshift, SNR, and inclination, rather than in it being an independent driver of classification. If the astrophysical mass distribution of hierarchical mergers evolves with redshift at a similar rate to the population assumed here, such evolution may be an \textit{astrophysical} selection effect against heavy hierarchical mergers, rather than a measurement selection effect. We also find that \LRp and \LRj classify binaries similarly across mass, indicating the inclusion or absence of \chieff as a classification dimension does not change the masses of binaries we are sensitive to identifying as hierarchical, even though \chieff measurement precision degrades with mass.

\textbf{\textit{Mass ratio}}: As with the grid injections, systems at the lowest mass ratios are more often successfully classified. Classified binaries with $q\lesssim0.4$ are twice as abundant as detectability alone would suggest. However, the impact in the overall mass ratio distribution is minor because of the surplus of signals near $q=0.5$. Due to number statistics, the highest SNR systems tend to be near the peak of the detected distribution, so the classified distribution remains centered near $q=0.5$, though skewed toward low $q$ vis-à-vis the detected distribution.

\textbf{\textit{Tilt}}: Systems that are nearly aligned or anti-aligned ($|\cos\tau_1|\approx1$) are not significantly precessing. Such systems are missed by the \chip-only metric: almost no binaries with $\cos\tau_1<-0.75$ are classified via \LRp. Aligned mergers are well-captured by the joint metric, comprising twice the fraction of \LRj-classified binaries than they do detected binaries. However, the most anti-aligned systems are not classified by the joint metric, either. This is explainable by the systematically larger measurement uncertainty for negative \chieff systems~\cite{Ng:2018neg}, in part due to their shorter inspirals. The detected-vs-classified distribution of \chieff shows comparable trends to that of $\cos\tau_1$ since \chieff is roughly proportional to $\cos\tau_1$ with $a_1\approx0.7$ and $a_2$ small in all cases.

\textbf{\textit{Precessing spin}}: It is unsurprising that classification via \chip alone monotonically improves with larger \chip. By incorporating \chieff information, the joint metric is nearly agnostic to the true \chip.

The inferred parameters of GW241011 and GW241110 are consistent with the detectable distributions of the six described parameters, but lie toward the edges of the \chip-classified distribution in tilt angle. As fewer than 1 in 4 simulated events were classified by \chip, the consistency between the observed events and simulated distributions, but that neither observed event achieved $\ln\mathrm{LR}>3$, is unsurprising. Given GW241011's high SNR and well-constrained nearly-aligned tilt, its infinite joint LR is consistent with expectation.

\subsubsection{Classification predictor}\label{sssec:logistic-regression}

We run a binary logistic regression on the CMC PE to estimate the probability of classification as a function of BBH parameters. While parameters have nonlinear and correlated effects on classification, we use this simple model to identify the dominant predictors while maintaining interpretability. We fit using redshift, cosine tilt, cosine inclination, source-frame primary mass, and mass ratio, reparameterizing tilt and inclination with their absolute values to account for the non-monotonic dependence of classification on these parameters. 

Using \LRp, redshift is the best predictor of classification, followed by inclination, tilt, and mass ratio. The information in redshift arises from its joint impact on SNR and inclination. As described in Sec.~\ref{sssec:cmc-merger-results}, while face-on inclination worsens precession measurement, it boosts SNR, such that the farthest detected systems are typically face-on and have less precise \chip measurements. We run residual tests to confirm this behavior, in which we linearly regress out a confounding variable then predict solely using the residual. Redshift, after removing its linear correlation with SNR, continues to classify well. Much of the remaining predictive power comes from inclination. The final bit of information in redshift appears to trace to source-frame mass: controlling for SNR, inclination, and mass eliminates all of redshift's predictive power. This may indicate an independent effect of mass on classification, but may also reflect the \textit{nonlinear} dependence of SNR on these variables that the linear residual model does not capture.

We verify these trends using a classifier trained on the grid PE. For those, mass ratio is the strongest predictor, followed by tilt and redshift. Both source- and detector-frame mass alone predict worse than chance. This supports that mass's mild effect in the CMC sample lay in its correlation with redshift and SNR. Mass ratio is a stronger predictor in the grid than the CMC mergers, likely because the CMC catalog produces few mergers below $q=0.3$ ($\sim2\%$ of detected mergers), limiting the range over which $q$'s effect can be assessed. These regression tests confirm that $q$ is the strongest intrinsic driver, while redshift and inclination are the extrinsic factors that most impact classification via effective spins.

\section{Conclusion}\label{sec:conclusion}

Measurement ability is a key aspect of characterizing populations. Motivated by the perhaps-unexpected low masses of the hierarchical merger candidates GW241011 and GW241110, we analyze the distribution of hierarchical mergers that can be classified as such. Assessing consistency between an event and a formation channel typically involves computing metrics, such as Bayes factors, from its inferred parameters. Since PE precision---especially for spin---is correlated with other event properties, some systems may not be identifiable as part of a population because their parameters cannot be sufficiently well-measured. Interpreting a classification outcome, particularly a null result, requires accounting not just for detectability but measurability. From a theoretical perspective, assessing whether a model makes testable predictions also entails understanding which regions of parameter space allow for confident classification.

Accounting for both selection effects and measurement capability in a simulated population of hierarchical mergers, we assessed classification via precession following Ref.~\cite{Payne:2024ywe}. We found unequal-mass mergers are more likely to be successfully classified, as are close, edge-on binaries. While measurements of \chieff worsen with primary mass as anticipated, measurements of component spin parameters and \chip are weakly affected. Primary mass therefore has minimal impact on hierarchical classification. Correlations between redshift, SNR, and inclination in realistic populations muddle these parameters' strength as sole predictors of precession measurement. We further introduce a classification metric in the joint space of \chieff and \chip that captures nearly-aligned and high-redshift mergers better than \chip alone.

Astrophysical interpretation of GW events is complicated by the fact that measurements have variable limitations. We offer one perspective with the classifiable distribution, which employs the fact that while measurement ability is variable, it is also predictable. As catalogs expand, placing apparent outliers in appropriate measurement context is key to avoiding over-interpretation~\cite{Fishbach:2019ckx, Mandel:2025qnh, Tenorio:2026dcc, Mould:2026nle}.

We reiterate that these classification metrics assume small natal spins of BHs. Population analyses suggest an extended \chip distribution~\cite{gwtc3pop, Banagiri:2025dmy, gwtc4pop, Plunkett:2026pxt, gwtc5pop}, which may be attributed to a subpopulation of hierarchical mergers, or to precessing first-generation mergers. High \chip may not unambiguously identify a hierarchical origin---and precession measurements may further be affected by waveform systematics and data quality~\cite{gwtc3cat, Payne:2022spz, Udall:2025bts, LIGOScientific:2025rsn, Ray:2025rtt}.

On the real data, we find four events are classified as hierarchical by the \chip-only metric. Assuming a classification success rate between 8 and 24\% (spanning those of Ref.~\cite{Payne:2024ywe} and here) with no false positives, this suggests $\sim6\text{--}19\%$ of observed mergers may be hierarchical. Recent analyses have identified an emerging subpopulation of rapidly spinning binaries from $\sim15\text{--}20\,M_\odot$, a feature not driven solely by GW241011 and GW241110~\cite{Antonini:2025ilj, Tong:2025xir, Plunkett:2026pxt, Alvarez-Lopez:2026ymo, Flanagan:2026ayy, gwtc5pop, OShaughnessy:2026dsr, OShaughnessy:2026xrd}, with an overall population fraction in line with the percent identified here. Rigorous tests of this subpopulation's association with hierarchical mergers, and understanding the events that comprise it, requires progress on multiple fronts beyond population modeling: improved stellar evolution models to better constrain first-generation spins, larger and higher-fidelity cluster simulations to map the parameter space of dynamical mergers~\cite[e.g.,][]{Mai:2025jmk, OConnor:2026lxl}, and refinement of single-event classification frameworks that acknowledge measurement capability. Together, these advances will bolster our ability to identify hierarchical mergers and reconstruct the environments that form and merge black holes.

\acknowledgments

We thank Sofía Álvarez-López, Tom Callister, Maya Fishbach, Beth Flanagan, Jack Heinzel, Simona Miller, Noah Wolfe, and Claire Ye for discussions.
C.P. is supported by the National Science Foundation Graduate Research Fellowship under Grant No. DGE-2141064.
M.Z. gratefully acknowledges funding from the Brinson Foundation in support of astrophysics research at the Adler Planetarium. 
S.V. is partially supported by the NSF grant no. PHY-2045740.
The authors are grateful for computational resources provided by the LIGO Laboratory supported by National Science Foundation Grants PHY-0757058 and PHY-0823459.
This material is based upon work supported by NSF's LIGO Laboratory which is a major facility fully funded by the National Science Foundation and has made use of data or software obtained from the Gravitational Wave Open Science Center (gwosc.org), a service of the LIGO Scientific Collaboration, the Virgo Collaboration, and KAGRA.

\appendix

\section{Detectability and injection details}
\label{app:detectability}

For the binaries drawn from the CMC catalog, we randomly generate the remaining binary parameters from standard distributions. We assume an isotropic distribution of spin-orbit misalignment angles, as anticipated for a cluster formation scenario. Although the CMC catalog assumes black holes are born with zero natal spin, we add small spins $\chi\sim\mathcal N(\mu=0,\sigma=0.1)$ to the 1G black holes as a more realistic scenario. While not entirely self-consistent, this should not significantly change the mass distribution of first-generation or hierarchical mergers, but it would decrease the rate of hierarchical mergers due to the higher natal kicks of remnants whose progenitors had spin. The rate should not affect conclusions about classification, which depends on the normalized detectable distribution and measurability.

We set the binary phase and polarization for the grid injections to zero. For all injections, we simulate the GW signals using the frequency-domain waveform \textsc{IMRPhenomXPHM} \cite{Pratten:2020ceb}. We run each system through a three-detector LIGO--Virgo network at design sensitivity for the fourth observing run~\cite{LIGO-T2000012} using zero-noise realizations to prevent variation due to noise fluctuations as opposed to signal properties.

For the CMC signals, we simulate a more realistic detection process using representative power spectral densities (PSDs) and Gaussian noise. For the LIGO Hanford and Livingston detectors, we use the published representative PSD for the first part of the fourth observing run \cite{LIGOScientific:2025hdt}; for the Virgo detector, we use the O3 representative PSD \cite{Virgo:2022ysc}, which is comparable to its sensitivity when it joined the second part of the fourth observing run. Finally, we impose a detection threshold of a network matched-filter SNR of 11. While this is strictly an unphysical selection criterion because it depends on the true signal parameters~\cite{Essick:2023upv}, the introduced bias is likely negligible at current sensitivities.

\begin{table}[!ht]
\begin{tabular}{ll}
\hline \hline
Parameter & Prior \\ \hline
$\mathcal M$ & $\mathcal U(\mathcal M_\mathrm{min},\, \mathcal M_\mathrm{max})$ \\
$q$ & $\mathcal U(0.1,1)$ \\
Spin magnitudes & Uniform \\
Spin tilts & Isotropic \\
$D_L$ & $\mathcal U(\frac14 D_L^\mathrm{true}, 4 D_L^\mathrm{true})$ \\
Zenith & Isotropic \\
Azimuth & Uniform \\
Polarization angle & Uniform \\
Inclination angle & Isotropic \\
Phase & Uniform \\
Coalescence time & $\mathcal U(t_c^\mathrm{true}-0.1, \, t_c^\mathrm{true}+0.1)$ \\
\hline \hline
\end{tabular}
\caption{Parameter estimation priors used during inference. ``Uniform" and ``Isotropic" priors are over the relevant domain. Otherwise, we use $\mathcal U$ to denote a uniform prior between the given bounds. We elaborate on $\mathcal M$ in the text.}
\label{tab:pepriors}
\end{table}

\section{Parameter estimation}
\label{app:PE}

We use the \textsc{Dynesty}~\cite{Speagle:2019ivv} sampler implemented in \textsc{Bilby}~\cite{bilby_paper} to draw posterior samples for every simulated system. Because these events are highly precessing, we use the full likelihood~\cite{Veitch:2014wba, bilby_paper} as opposed to a heterodyned approximation~\cite{Cornish:2010kf, Zackay:2018qdy, Cornish:2021lje}. While the full likelihood increases computational cost, we found it was necessary for accurate PE: for a subset of signals analyzed with the relative binning likelihood, the posteriors were visually different and the reweighting efficiencies to the full likelihood were poor.

We use standard priors~\cite{Romero-Shaw:2020owr}, summarized in Tab.~\ref{tab:pepriors}. For chirp mass $\mathcal M$, we heuristically set the prior width in a manner similar to Ref.~\cite{Wolfe:2025yxu}. We set $\Delta\mathcal M = 10^{-2}(\mathcal M^\mathrm{true})^{5/2}$ and find the lowest allowed chirp mass consistent with a signal of the given length, $\mathcal M_\mathrm{low}$. We then select $\mathcal M_\mathrm{min} = \max\left(\mathcal M^\mathrm{true} - \frac12\Delta \mathcal M, \, \mathcal M_\mathrm{low}\right)$, and $\mathcal M_\mathrm{max} = \min\left(\mathcal M^\mathrm{true} + \frac12\Delta \mathcal M, \, 250 M_\odot \right)$. We sample $\mathcal M$ in the detector frame and coalescence time $t_c$, zenith, and azimuth with respect to LIGO Hanford. We set the reference and minimum frequencies to 20 Hz.

\section{Additional classification metric results}\label{app:metrics}

\begin{figure}[!htb]
\centering
\includegraphics[width=\columnwidth]{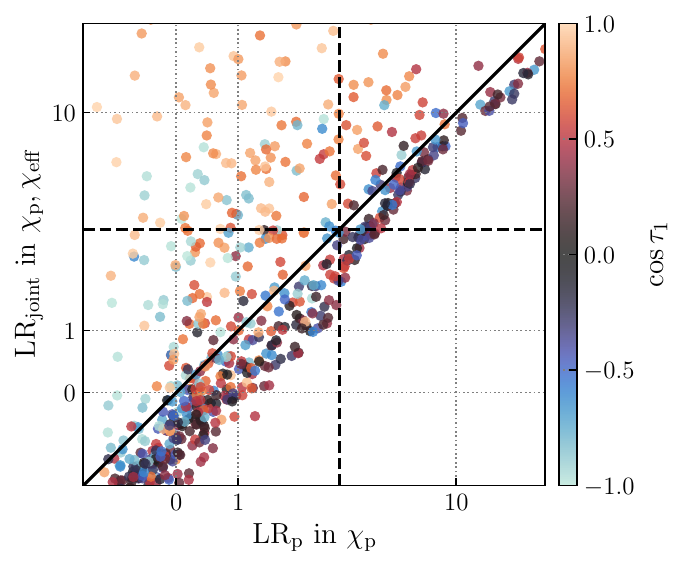}
\caption{Comparison of the log likelihood ratio statistic in \chip alone to that in the joint space of \chieff and \chip for the simulated mergers from the CMC catalog. Mergers are colored by cosine tilt angle. The solid black line indicates equality, while dashed lines mark the classification limits of $\ln\mathrm{LR}=3$.}
\label{fig:compare-metrics}
\end{figure}

We compare the \chip-only and joint \chieff--\chip metrics for the CMC injections in Fig.~\ref{fig:compare-metrics}. The solid black line indicates equality, while the dashed lines at $\ln\mathrm{LR}=3$ separate classified and unclassified binaries. Some binaries lie beyond the plot limits; 52 have either ln LR $<-1.5$ and 38 have either ln LR $> 25$. Most mergers have \LRj below their \LRp due to the larger prior volume in the joint space, which downweights the joint metric. The upper left quadrant contains mergers missed by \chip alone but classified by the joint metric. As also seen in Figs.~\ref{fig:cmc-pe-det-vs-class} and~\ref{fig:cmc-pe-frac-change}, these tend to be the nearly aligned mergers. The lower right quadrant contains 23 events classified by \chip alone but missed by the joint metric. There is no strong trend in their parameters, other than that they are typically in-plane (expected due to number statistics) and typically marginal classifications: of events found by \chip but missed by the joint metric, the highest $\ln\mathrm{LR_p}$ is 4.4. 

\begin{figure}[!htb]
\centering
\includegraphics[width=\columnwidth]{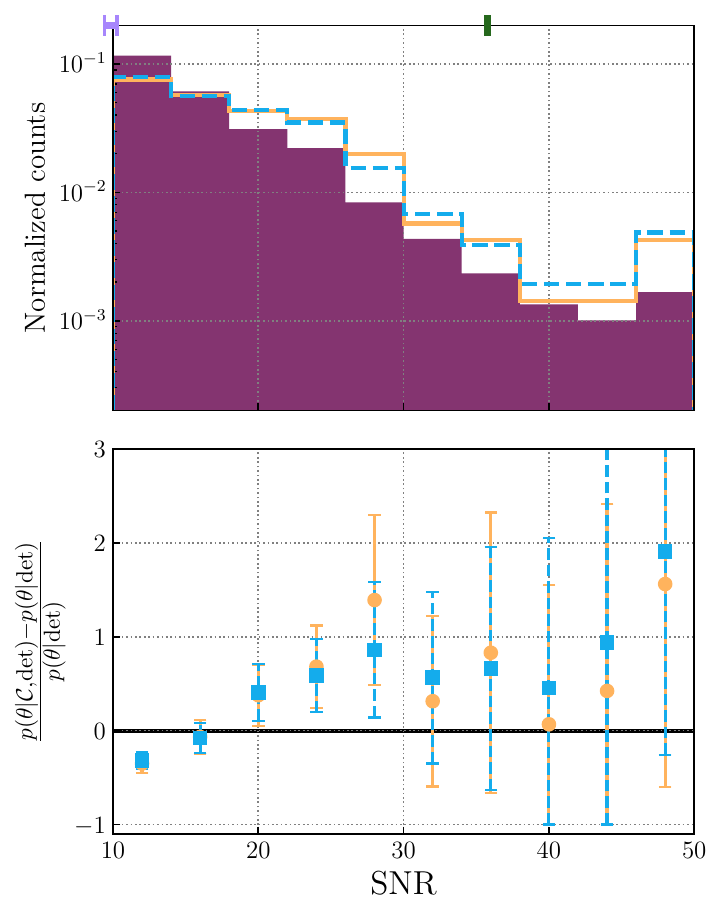}
\caption{As in Figs.~\ref{fig:cmc-pe-det-vs-class} and~\ref{fig:cmc-pe-frac-change}, but for SNR.}
\label{fig:cmc-pe-snr}
\end{figure}

Fig.~\ref{fig:cmc-pe-snr} shows the SNR dependence of classification for both metrics.  There is a clear increasing trend in classification rate up to SNR$\sim30$, but the trend is not strictly monotonic at higher SNRs. This is due to a combination of low number statistics (and resulting uncertainty in classification rate) and the inclination dependence: the highest-SNR signals tend to be face-on and have worse-measured precession. SNR is a strong predictor of overall measurement precision. However, in realistic populations, SNR's correlation with inclination reduces its power as a sole predictor of precession measurement.

\begin{figure*}[!htb]
\centering
\includegraphics[width=0.85\textwidth]{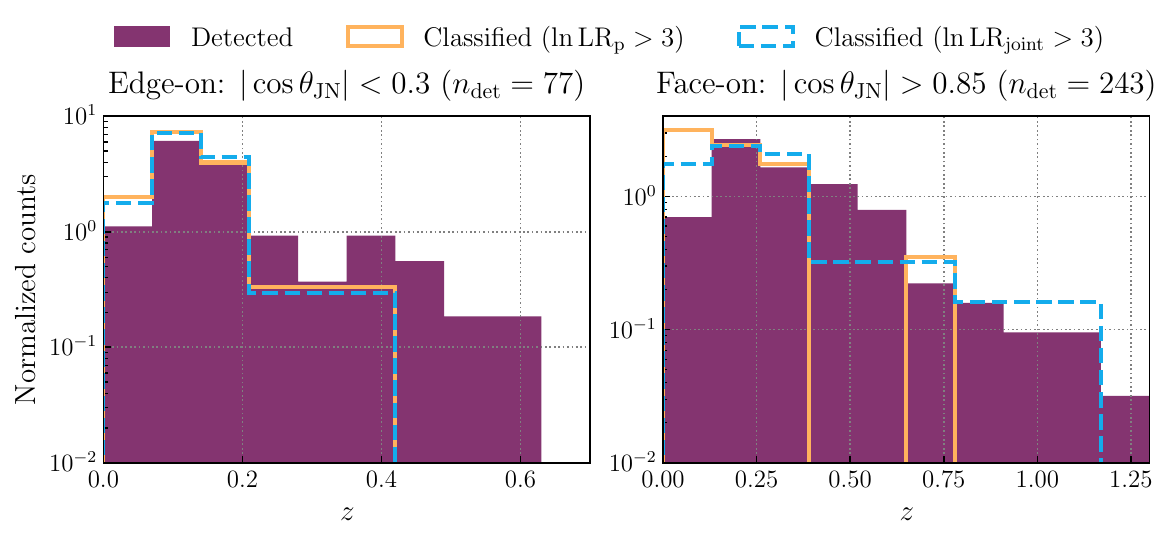}
\caption{Comparison of the redshift dependence of the \chip-only and joint classification metrics for two bins in inclination angle $\theta_\mathrm{JN}$. The left panel shows nearly ``edge-on" mergers ($|\cos\theta_\mathrm{JN}|<0.3$, within $\sim18^\circ$ of edge-on), while the right panel shows ``face-on" mergers ($|\cos\theta_\mathrm{JN}|>0.85$, within $\sim30^\circ$ of face-on).}
\label{fig:compare-redshift-dependence}
\end{figure*}

We also investigate the shallower redshift dependence of the joint metric identified in Sec.~\ref{sssec:cmc-merger-results}. Fig.~\ref{fig:compare-redshift-dependence} shows the detected and classified distributions over redshift as in Fig.~\ref{fig:cmc-pe-det-vs-class}, further stratified by inclination. Nearly edge-on events are detected less often and at lower redshifts due to their lower SNRs, but allow for better precession measurement. Edge-on events are classified at higher rates by both metrics. The joint and \chip-only metrics have nearly identical redshift dependence for edge-on events, but the joint metric classifies face-on events to much higher redshifts.  This supports the notion that the difference in the metrics' overall redshift dependence comes from events whose lack of inclination inhibits precession measurement, but \chieff can still be well-measured.

\subsection{Classification in component parameters}

The intrinsic parameters of $q$, $a_1$ and $\tau_1$ may offer more direct discriminatory power than effective parameters. We explore the likelihood ratios in these parameters and their joint distribution. For our injected events, the PE priors on these parameters are uniform, so the likelihood ratio is a traditional posterior Bayes factor on the hypotheses of being inside versus outside a chosen volume $\mathcal V$. For GW241011 and GW241110, the PE prior is uniform in components, not $q$, so we again divide out the prior dependence. The thresholds in the parameters and LR must be informed by an assumed distribution of first-generation mergers. We set heuristic parameter limits to explore how these LRs change across parameter space but refrain from claiming a particular limit as the appropriate threshold for classification.

\begin{figure*}[!htbp]
\centering
\includegraphics[width=0.9\textwidth]{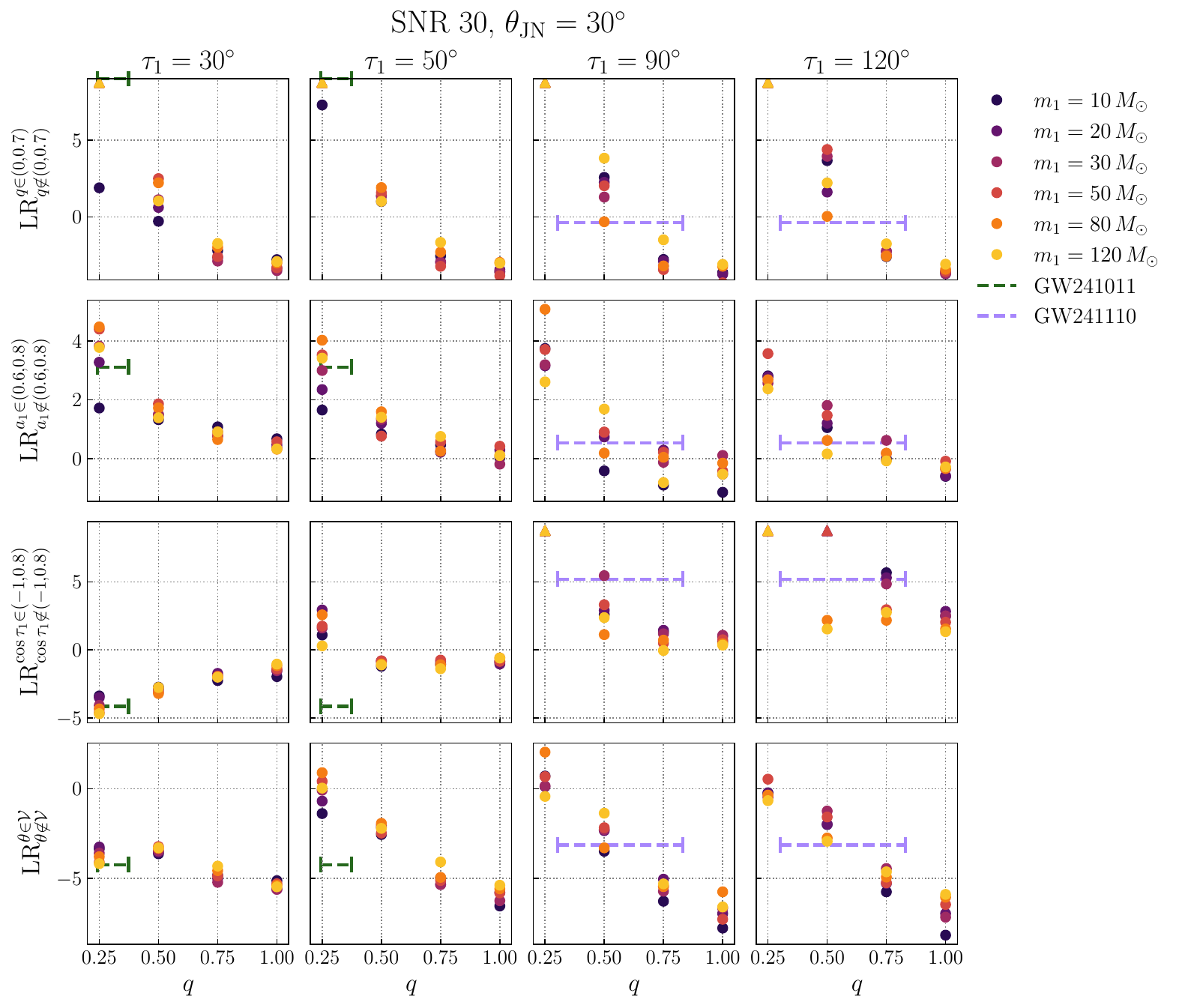}
\caption{As in Fig.~\ref{fig:PE_grid_LR}, but for the constituent parameters $q,\, a_1,\,\cos\tau_1$, and their joint distribution for the SNR 30 injections.}
\label{fig:PE_grid_PR_snr_30}
\end{figure*}

The component LRs following many of the trends of the \chip and \chieff--\chip LRs. Fig.~\ref{fig:PE_grid_PR_snr_30} shows the individual component and joint-component LRs for the SNR 30 grid injections. As in Fig.~\ref{fig:PE_grid_LR}, we plot LR as a function of $q$, with color corresponding to primary mass. The first three rows plot the LRs for $q$, $a_1$, and $\tau_1$ individually, while the final row shows the statistic in the joint space.

Mass ratio influences some component spin parameter measurements. It has a significant impact on spin magnitude, with the lowest mass ratios systematically producing higher LRs. Tilt measurements improve with lower $q$, but its effect on LR is obfuscated by the interplay between measurement precision and the large prior coverage of the chosen threshold. At $\tau_1=30^\circ$, LR in $\cos\tau_1$ \textit{increases} with more equal mass ratio. This is a spurious effect of the chosen threshold ($\cos\tau_1<0.8$) covering 90\% of the prior width, while the true $\cos\tau_1\approx0.86$. At $q=1$, tilts are measured more poorly, cover a larger fraction of the prior, and produce higher LRs. At $\tau_1 = 50^\circ$, these effects roughly cancel such that LR is flat across $q$, while at higher tilts it improves with lower $q$.

\begin{figure*}[!htb]
\centering
\includegraphics[width=0.8\textwidth]{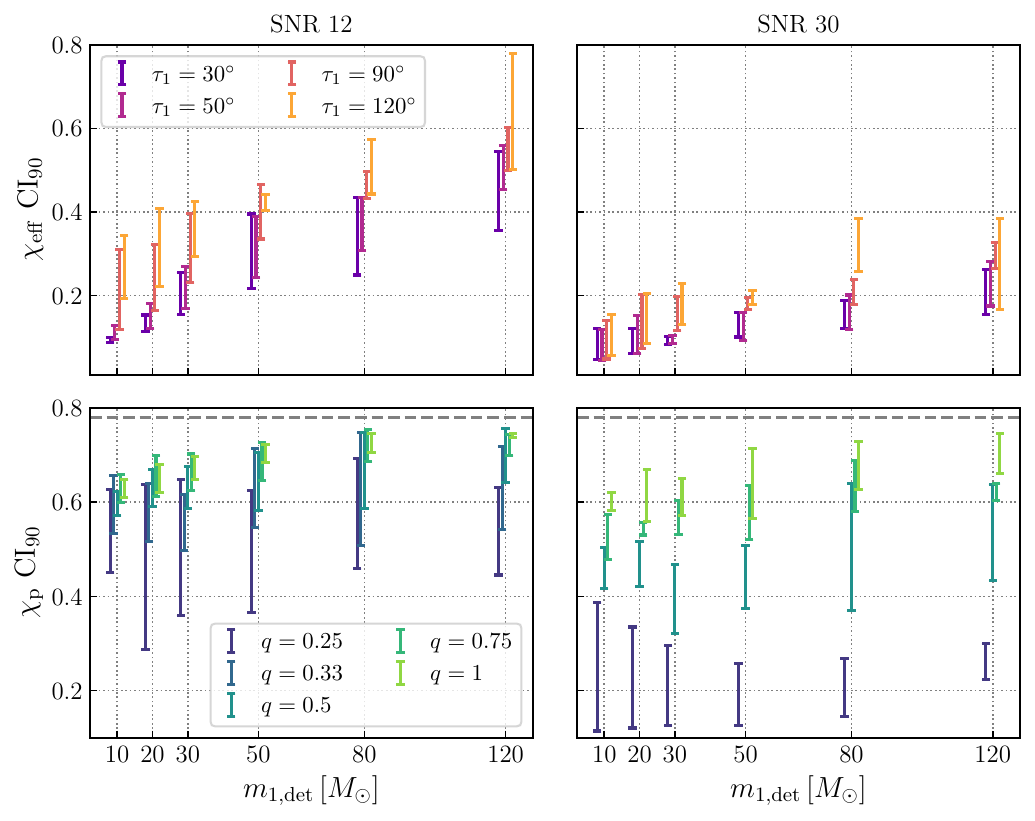}
\caption{The posterior 90\% credible widths for \chieff (top) and \chip bottom for the grid injections at SNR 12 (left) and 30 (right) at $\theta_\mathrm{JN}=30^\circ$ as a function of mass. The injections are further stratified by tilt angle for \chieff and mass ratio for \chip; these are shown in different colors with slight offsets from the nominal primary mass. The bar spans the range of widths across the remaining varied parameter, which is mass ratio for \chieff and tilt for \chip. The gray dashed line on the lower panels marks the 90\% prior width for \chip.}
\label{fig:PE-widths}
\end{figure*}

\section{Measurement behavior across mass}
\label{app:mass}

We illustrate the mass-dependent behavior of the \chieff and \chip posteriors in Fig.~\ref{fig:PE-widths}. The left (right) column is for SNR 12 (30); all posteriors are for binaries at $\theta_\mathrm{JN}=30^\circ$. The top row shows the 90\% widths of the \chieff posteriors at each primary mass. We further bin by tilt angle, which has the second largest effect, and plot the spread of the widths across different mass ratios with an error bar. The lower panels depict the posterior widths for \chip, secondarily binned by mass ratio, such that the error bars encapsulate the spread of posterior widths over tilt angle at the given mass and mass ratio.

The \chieff posterior widths increase with primary mass due to fewer cycles in-band, as anticipated. At a given mass, a more misaligned tilts produces a more poorly measured \chieff due to the shorter inspiral~\cite{Ng:2018neg}. The \chip posterior widths are instead most correlated with mass ratio. The same misalignment at more unequal masses leads to more observable precession. While mass has minimal effect on the width of the \chip posterior for $q\in\{0.25,\,0.33,\,0.5\}$, there are slight broadening trends for $q=0.75$ and $q=1$. This suggests a mild mass dependence that is tempered at unequal mass ratios.

The mass dependence at equal masses likely comes from the loss of precession cycles, similar to how fewer inspiral cycles degrade \chieff measurement. Ref.~\cite{Green:2020ptm} find precessing SNR decreases with mass, and thus expect broader \chip posteriors, though the effect is weaker than for \chieff. We find the mass dependence disappears at unequal mass ratios, which may arise from an interplay between precession, mass ratio, and mass with higher order modes (HOMs). Any HOM effect could not have been predicted by Ref.~\cite{Green:2020ptm}, who simulated signals using a waveform that lacked higher harmonics. At $q=1$, the waveform is dominated by the standard (2, 2) mode of gravitational radiation, while HOMs are excited at unequal masses. HOMs in the merger possibly encode precession information that counters the mass dependence. A full exploration of the waveform phenomenology driving this effect would require substantially more simulations and is beyond the scope of this work.

We also note that some spin parameters, including \chip, are time-dependent in the presence of precession and must be defined at a chosen reference frequency. For systems at different masses, the same reference frequency corresponds to a different number of cycles before merger, and therefore the quantity under consideration is not identical across mass. The shown behavior of \chip across mass instead indicates the outcome of a typical analysis process, in which the reference frequency is often chosen to be 20 Hz for all events~\cite{LIGOScientific:2025zdk}. Spin vectors may instead be back-evolved to an infinite reference distance~\cite{Mould:2021xst}.

\bibliography{paper}

\end{document}